\documentstyle[12pt]{article}
\begin{document}
\begin{center}
Temporal chaos in discrete one dimensional gravity model of traffic flow\\
        Elman Mohammed Shahverdiev,\footnote{e-mail:shahverdiev@lan.ab.az\\
On leave from Institute of Physics, 33, H.Javid avenue, Baku 370143, 
Azerbaijan}\\
Department of Information Science,\footnote{e-mail:elman@ai.is.saga-u.ac.jp}
Saga University, Saga 840, Japan\\
Shin-ichi Tadaki\\
Department of Information Science,\footnote{e-mail:tadaki@ai.is.saga-u.ac.jp}
Saga University, Saga 840, Japan\\
\end{center}
There are mainly two approaches to traffic flow dynamics: At a microscopic 
level, the system can be described in terms of variables such as the position 
and velocity of each vehicle (optimal velocity 
model [1,2 ],cellular automaton model [3,4 ]);at a macroscopic level important variables include the car 
density, average velocity, the rate of traffic flow, the total number of 
trips between two zones.(mean field theory [5-11], origin-destination (or the so-
called gravity)model [12,13]).The gravity model originates from an anology withNewton's gravitational law [13].
As a rule traffic flow dynamics is the nonlinear one.Nowadays it is 
well-known that some deterministic nonlinear dynamical systems depending on 
the value of system's  parameters exhibit unpredictable,chaotic behavior,see,e.g.[14-16] and references therein.
The main reason for such a behavior is the instability of the nonlinear 
system. Such an instability in general was undesirable not only in traffic 
flow dynamics ,but also in dynamical systems in mechanics, engineering,etc, 
due to the frightening nature of unpredictability as the unstability 
could lead to chaos. The stability or unstability of traffic flow dynamics
are highly valued conceptions in traffic management and planning.\\ 
Since the pioneering papers [17,18 ] on chaos control theory,the attitude to chaos 
has been changed dramatically. Nowadays in some situations chaotic behavior 
is considered even as an advantage.In general the main idea behind chaos 
control theory is to modify the nonlinear systems' dynamics so that 
previously unstable states (fixed points,periodic states,etc.)now become 
stable.In practice such modifications could be realized by changing system's 
parameters,through some feedback or nonfeedback mechanisms,or even by 
changing the dynamical variables of the system in an 
``appropriate'' manner in ``due''time (adaptively or nonadaptively),etc.   
The interest to the chaos control theory is due to the
application of this phenomenen in secure communication, in modelling
of brain activity  and recognition processes, etc,.
Also methods of chaos control may result in the improved performance of 
chaotic systems.(For the latest comprehensive review of chaos and its control
see Focus Issue [19] and references therein;also see [20-22]).\\
In this Brief Report we report on the possible
chaotic behaivor in one dimensional gravity model of traffic flow dynamics.
The gravity model assumes that the number of trips between zones 
(origins and destinations) depends on the number produced at and attracted 
to each zone,and on the travel cost between zones.In the dynamic formulation 
of the gravity model the travel costs are a function of the number of trips 
between zones.According to [12],the discrete dynamic trip distribution  gravity model takes the form
$$x_{ij}(t+1)=f(c_{ij}(t)),\hspace*{3cm}(1)$$
where $x_{ij}$ is the relative number of trips from zone $i$ to zone $j$,normalised so that $\sum_{ij} x_{ij}=1$, $c_{ij}$ is the travel cost from zone $i$ to zone $j$
given the trips $x_{ij}$.\\
$$c_{ij}(t)=c_{ij}^{0}(1+\alpha (\frac{x_{ij}}{z_{ij}})^{\gamma}),\hspace*{1cm}(2)$$
where $c_{ij}^{0}$ is the uncongested travel cost,$q_{ij}$ is the relative
capacity of the roads between origin and destination and $\alpha$ and 
$\gamma$ are constants.
$f(c_{ij})$ is a function which relates the number of trips to the travel 
costs.The following cost function
$$f(c_{ij})=c_{ij}^{\mu}\exp(-\beta c_{ij}),\hspace*{3cm}(3)$$
where $\mu$ and $\beta$ are constants is refered to as combined cost 
function and unites  both the power and exponential forms of cost 
functions.
It is known that for continuous dynamical systems for chaotic behavior the 
number of dynamical variables should be three or more than that.For discrete
systems chaos is possible even in one dimensional systems.
According to [12],for the unconstrained 
(the model is the unconstrained one in the sense that
it cannot guarantee that the number of trips originating from or terminating at a given zone has a value which is predetermined.)and singly-constrained
(the model is the singly-constrained one in the sense that
it supposes that either 
the number of trips originating from or terminating at a given zone has a value which is predetermined.)
gravity models the dynamics of trip distribution model in the one dimensional
case could be written as
$$x(t+1)= A f(x(t))=A(c^{(0)})^{\mu}(1+\alpha (\frac{x(t)}{q})^{\gamma})^{\mu}\exp(-\beta c^{(0)}(1+\alpha (\frac {x(t)}{q})^{\gamma})),\hspace*{1cm}(4)$$
where $A$ is the normalizing constant factor;the definition of other 
constants are given above. 
Authors of [12] claim that one dimensionsal gravity model does not exhibit chaotic
behavior. We will show that this model could be reduced to one dimensional 
chaotic model known as exponential map in ecology [23]:
$$x(n+1)=f(x(n))=x(n)\exp(r(1-x(n))),\hspace*{4cm}(5)$$
where $r$ is positive control parameter of the chaotic mapping (5).\\
Indeed let us take $\mu =\gamma =1$. Then the mapping (4) could be written
in the following form:
$$x(t+1)=m_{1}(1+mx(t))\exp(-\beta c^{(0)}mx(t)),\hspace*{5cm}(6)$$
where $m_{1}=Ac^{(0)}, m=\frac{\alpha}{q}$.Further by 
linear transformation of variables $y=1+mx(t)$ the mapping (6) could be 
related to the mapping:
$$y(t+1)=m_{2}y(t)\exp(\beta c^{(0)}(1-y(t))),\hspace*{4cm}(7)$$
where $m_{2}=m_{1}\exp (-\beta c^{(0)})$. Comparing (5) and (7) one can see that in the 
gravity model the $\beta c^{(0)}$ could be taken as a control parameter.
Thus we have shown that the temporal chaotic behavior is possible even in
 discrete one 
dimensional gravity model for traffic flow dynamics.Moreover this chaotic behavor could be controlled by the constant rate harvesting approach developed in 
[24] for unimodal one dimensional mappings,including (5) or by some other methods for one dimensional dynamical systems.\\
Acknowledgments\\
The author thanks the JSPS for the Fellowship.\\

\end{document}